\documentclass[dvips,, 12pt]{article}
\usepackage{latexsym, epsfig, amsmath, graphicx, amssymb}
\usepackage{mathrsfs, amsfonts, graphics}
\begin{document}

\begin{center}
\textbf{\Large Asymptotic Behaviour of the Proper Length and Volume of the Schwarzschild
Singularity}\\[1.2cm]
Asghar Qadir$^{1}$ and Azad A. Siddiqui$^{2,3}$\\[3ex]
$^{1}$Center for Advanced Mathematics and Physics, Campus of EME College,
National University of Science and Technology, Peshawar Road, Rawalpindi,
Pakistan; and Department of Mathematical Sciences, King Fahd University of
Petroleum and Minerals, Dhahran, 31261, E-mail: aqadirs@comsats.net.pk
\\[0pt]
$^{2}$Department of Basic Science and Humanities, EME College, National University
of Science and Technology, Peshawar Road, Rawalpindi, Pakistan,\\[0pt]
E-mail: azad@ceme.edu.pk \\[0pt]
$^{3}$\emph{Present Address}: Department of Physics and Measurement Technology,
Linköping University, SE-583 81 Linköping, Sweden, E-mail: azad@ifm.liu.se

\bigskip

\bigskip

\bigskip

\textbf{Abstract}
\end{center}

\begin{quotation}
Though popular presentations give the Schwarzschild singularity as a point
it is known that it is spacelike and not timelike. Thus it has a ``length''
and is not a ``point''. In fact, its length must necessarily be infinite. It
has been proved that the proper length of the Qadir-Wheeler suture model
goes to infinity [1], while its proper volume shrinks to zero, and the
asymptotic behaviour of the length and volume have been calculated. That
model consists of two Friedmann sections connected by a Schwarzschild
``suture''. The question arises whether a similar analysis could provide the
asymptotic behaviour of the Schwarzschild black hole near the singularity.
It is proved here that, unlike the behaviour for the suture model, for the
Schwarzschild essential singularity $\Delta s$ $\thicksim $ $K^{1/3}\ln K$
and $V\thicksim $ $K^{-1}\ln K$, where $K$ is the mean extrinsic curvature,
or the York time.
\end{quotation}

\section{Introduction}

Despite popular presentations, it is well known [2] that the Schwarzschild
(curvature) singularity is spacelike and hence is \textit{not} a point.
However, some of the implications of this fact have not been explicitly
brought out and considered. If it is in effect, a line, it is natural to ask
what its length is? The answer is intuitively obvious. Since it is singular
it must have an infinite or a zero length. Further, since the singularity is
not a point the length must be infinite. This answer raises two new
questions. First, how is there a ``collapse'' if the length has gone
infinite? Second, how does the length approach infinity as the singularity
is approached? The answer to the former question is reasonably easy to see.
It must be that the \textit{volume} approaches zero even as the \textit{
length} approaches infinity. The second can not be answered without first
agreeing on the choice of time slicing for the spacetime. Complete foliation
of the Schwarzschild spacetime by spacelike hypersurfaces of constant mean
extrinsic curvature, York or $K$-slices, had been obtained earlier [3] and
then proved to be complete [4]. This paper will deal with these questions in
more detail and particularly with the second question.

The question raised above had arisen earlier in the context of the ``suture
model'' [5, 6] which modeled the formation of a black hole in a closed
universe. This was done by attaching a slice of a denser closed Friedmann
universe model to a rarer closed Friedmann model with a slice of equal mass
cut out. Since the two sections evolve at different rates a matching at one
instant will not be able to hold for any other instant. To keep them
together, they are attached by a ``suture'' with a Schwarzschild geometry.
It was required that the suture disappear at the big bang and evolve to
``stitch'' the two Friedmann sections together. Thus the denser Friedmann
section could be regarded as a density perturbation in the rarer Friedmann
section. At some stage the denser section collapses below the Schwarzschild
radius, as seen from outside, and can therefore be regarded as a black hole.
This model was sliced using the York time [2], i.e. by spacelike
hypersurfaces of constant mean extrinsic curvature, $K$. The essential
finding was that the black hole and big crunch singularities form
simultaneously according to this slicing.

In the process of foliating the suture model, it was found that enormous
amounts of computer time were required for obtaining the time-slices as one
approached the big crunch. Since the Friedmann sections had collapsed to
very small sizes, it would appear that it was the Schwarzschild suture that
was causing the problem. However, there had been no clear understanding
\textit{why} this was happening. Subsequently, a heuristic argument was
given which showed that the proper length of the hypersurface tends to
infinity as the mean extrinsic curvature tends to zero [7]. This raised the
question of whether there was, in fact, a collapse. By the same heuristic
argument it was found that the volume of the slice does go to zero as
required for collapse. A more rigorous analysis, by Hussain and Qadir (HQ)
[1] showed that the basic expectations of the heuristic argument, that the
proper length diverges and the proper volume goes to zero, are correct (but
the explicit form deduced heuristically turned out to be incorrect).

A naive analysis already shows the divergence of the proper length. The
Schwarzschild metric written in the usual Schwarzschild coordinates,

\begin{equation}
ds^{2}=-(1-r_{s}/r)dt^{2}+(1-r_{s}/r)^{-1}dr^{2}+r^{2}d\Omega ^{2},
\label{a}
\end{equation}
is only valid for $r>r_{s}=2Gm/c^{2}$. However, if we naively apply it for $
r<r_{s}$ the $dt^{2}$ term gets a negative sign and the $dr^{2}$ term a
positive sign. Thus $r$ becomes a time parameter and $t$ a space parameter.
Thus, a time slice would be one in which we put $r$ equal to a constant.
Taking the slice at a given value of $\theta $ and $\varphi $ we see that
the proper length becomes

\begin{equation}
\Delta s=\surd \overline{(r_{s}/r-1)}\int dt,  \label{b}
\end{equation}
over some appropriate range of integration. Thus $\Delta s\thicksim r^{-1/2}$
. The problem is that the range of integration looks as if it should, itself
be infinite. Further, if the range of integration were ignored, the volume, $
V$, appears to go as $r^{3/2}$ and hence tends to $0$. However, the range of
integration casts doubt on whether the collapse genuinely occurs. The $0$
may have to be multiplied by an $\infty $ to give a finite (or even
infinite) value. As such, we need to calculate the behaviour more carefully.

In this paper we will demonstrate that the asymptotic behaviour of the
Schwarzschild singularity follows the expectations arising from the suture
model and the naive arguments given above. We will first perform the
calculations in the compactified Kruskal Szekres (CKS) coordinates [2, 8],
used for constructing the Carter-Penrose (CP) diagram and then use the
K-slice [3, 4, 9, 10] behaviour. We conclude with a summary and discussion.

\section{Asymptotic Behaviour of Length of the Singularity in CKS Coordinates
}

The Schwarzschild metric in compactified Kruskal Szekres (CKS) coordinates $
(\psi ,\xi )$ is given as [2, 8]
\begin{equation}
ds^{2}=f^{2}(r)\frac{-d\psi ^{2}+d\xi ^{2}}{4\left( \cos ^{2}\frac{\psi }{2}
-\sin ^{2}\frac{\xi }{2}\right) ^{2}}+r^{2}d\Omega ^{2},  \label{1}
\end{equation}
where $-\pi /2<\psi <\pi /2$ and $-\pi <\xi <\pi $,

\begin{equation}
f^{2}(r)=\frac{4r_{s}^{3}}{r}e^{-\frac{r}{2m}},  \label{2}
\end{equation}
and the radial parameter $r$ and the CKS coordinates $(\psi ,\xi )$ are
related by
\begin{equation}
\left( 1-\frac{r}{r_{s}}\right) \exp \left( \frac{r}{r_{s}}\right) =\tan
\left( \frac{\psi +\xi }{2}\right) \tan \left( \frac{\psi -\xi }{2}\right) .
\label{3a}
\end{equation}

To determine the proper length on the hypersurface, $\Delta s$,
\begin{equation}
\Delta s=\int_{s_{1}}^{s_{2}}ds=\int_{s_{1}}^{s_{2}}\frac{f(r)}{2\left( \cos
^{2}\frac{\psi }{2}-\sin ^{2}\frac{\xi }{2}\right) }\left[ -\left( \frac{
d\psi }{ds}\right) ^{2}+\left( \frac{d\xi }{ds}\right) ^{2}\right] ^{\frac{1
}{2}}ds,  \label{3}
\end{equation}
where the $\psi $ and $\xi $ lie on the chosen hypersurface. Even though
there is no geometrical significance to it, let us first try taking $\psi
=constt.$ on the hypersurface, for the sake of simplicity. Consider the
hypersurface close to $\psi =\pi /2$. Thus, putting $\psi =\pi /2-\epsilon $
(i.e. $r\rightarrow 0$), Eq.$\left( \ref{3}\right) $becomes

\begin{equation}
\Delta s=\int\limits_{-\pi /2-\epsilon }^{\pi /2+\epsilon }\frac{f(r)}{\sin
\epsilon +\cos \xi }d\xi ,  \label{4}
\end{equation}
and $f^{2}(r)$ can be given as

\begin{equation}
f^{2}(r)=4c^{2}r_{s}^{2}\epsilon ^{-1/4}\left[ 1+\epsilon ^{1/2}+O\left(
\epsilon \right) \right] ,  \label{5}
\end{equation}
where $c$ is a constant.

Using Eq.$\left( \ref{5}\right) $ in Eq.$\left( \ref{4}\right) $ and
integrating we obtain

\begin{equation}
\Delta s=2cr_{s}\epsilon ^{-1/8}\ln \left( \frac{2}{\epsilon }\right) \left[
1+\frac{1}{2}\epsilon ^{1/2}+O\left( \epsilon \right) \right] .  \label{6}
\end{equation}
Eq.$\left( \ref{6}\right) $ shows that $\Delta s\thicksim -\epsilon
^{-1/8}\ln \left( \epsilon \right) $. The constant, $c$, hides a problem. It
depends on the spacelike coordinate $\xi $, in a secant function, which
diverges as $\xi \rightarrow \pm \pi /2$. Thus the length has an extra
infinite factor multiplying it. This problem can be resolved by going to the
$K$, or York, slices, which have the added advantage of geometrical
significance.

\section{Asymptotic Behaviour of the Length of the Singularity Along the
York Slice}

The proper length of the singularity along the hypersurface of simultaneity,
defined by York time, can be obtained from the analysis of HQ. The only
modification to be made is that they used the boundaries of the
Schwarzschild region in the suture model, so that on a given slice they took
the spacelike coordinate $\xi $ to lie between certain given $\xi _{1}$ and $
\xi _{2}$. For our present purpose $-\pi /2\leqslant \xi \leqslant \pi /2$
at $r=0$. HQ found that

\begin{equation}
\Delta s=3^{5/6}\sqrt{2}r_{s}\ln \frac{\tan \xi _{2}+\sec \xi _{2}}{\tan \xi
_{1}+\sec \xi _{1}}(Kr_{s})^{1/3}\times \left[ 1+\frac{5}{4}
3^{-1/3}(Kr_{s})^{-2/3}+O(Kr_{s})^{-4/3}\right] .  \label{7}
\end{equation}
Since $\sec \xi $ becomes infinite at $\pm \pi /2$, at first sight it
appears that we need to take the limit as we approach these points. To this
end we take $\xi _{2}=\psi -2\delta $, with $\psi =\pi /2-\epsilon $, where $
\delta $ and $\epsilon $ are small. Using these values in Eq.$\left( \ref{3a}
\right) $, in order to reach $\psi =\pi /2$ in the limits $\delta ,$ $
\epsilon \rightarrow 0$, it is required that $\delta \geq \epsilon $ ( for $
\delta <\epsilon $ one reaches to the line $\psi =\xi $ first, which creates
problems in reaching the singularity). It has been shown by HQ that $
\epsilon $ goes to zero as $r^{2}$ and $r$ goes to zero as $K^{-2/3}$,
therefore $\epsilon $ goes to zero as $K^{-4/3}$.

Now taking $\xi _{1}=0$ and $\xi _{2}$ as mentioned above, and using the
extreme value of $\delta $ i.e. $\delta =\epsilon $ in Eq.$\left( \ref{7}
\right) $, we obtain

\begin{equation}
\Delta s=3^{-1/6}8\sqrt{2}r_{s}(Kr_{s})^{1/3}\ln \left( Kr_{s}\right) \times
\left[ 1+\frac{5}{4}3^{-1/3}(Kr_{s})^{-2/3}+O(Kr_{s})^{-4/3}\right] .
\label{7a}
\end{equation}
The volume, $V$, is given by

\begin{equation}
V=3^{-1/6}8\sqrt{2}\pi r_{s}\left( Kr_{s}\right) ^{-1}\ln \left(
Kr_{s}\right) \times \left[ 1+\frac{5}{4}%
3^{-1/3}(Kr_{s})^{-2/3}+O(Kr_{s})^{-4/3}\right] .  \label{7b}
\end{equation}

Thus the basic results of HQ can be seen to hold here as well, but the
proper length of the singularity goes to infinity even faster as there is an
extra factor, $\ln \left( Kr_{s}\right) $, in Eq.$\left( \ref{7a}\right) $.
The significance of this factor is discussed in the conclusion. The second
term in the brackets in Eqs.$\left( \ref{7a}\right) $ and $\left( \ref{7b}
\right) $ becomes less than $1$ for $Kr_{s}>0.8$. Therefore the asymptotic
behaviour given above is applicable for values of $Kr_{s}$ much greater than
$0.8$. In the following section we present a foliation by $K$-slices upto
this limit. There are problems of ill conditioning (excessive sensitivity)
to initial values above $0.5$, because of which we have not carried the
calculation beyond the bare $0.8$.

\section{Foliation of the Schwarzschild spacetime by K-slices}

As mentioned in the introduction, a complete $K$-slicing, had been obtained
earlier. In this foliation it was required that the foliating hypersurfaces
have zero slope at $\xi =0$ and go to spacelike infinity. In order to find
the length of the singularity, a more appropriate foliation would be one
which stays inside the event horizon and in the limit coincides with the
singularity at $r=0$. For this purpose we require that the foliating
hypersurfaces have zero slope at $\xi =0$ and $\psi =(1+\sigma )\xi $ ($
\sigma <<1$). Using a procedure similar to that adopted in [3, 4] we have
obtain a foliation upto $Kr_{s}=0.8$ for $\sigma =0.1$ (see figure 1). Table
1 gives the length and the maximum height of the hypersurfaces for different
values of $Kr_{s}$.
\begin{figure}[htb]
  \begin{center}
  \includegraphics[width=1.0\textwidth]{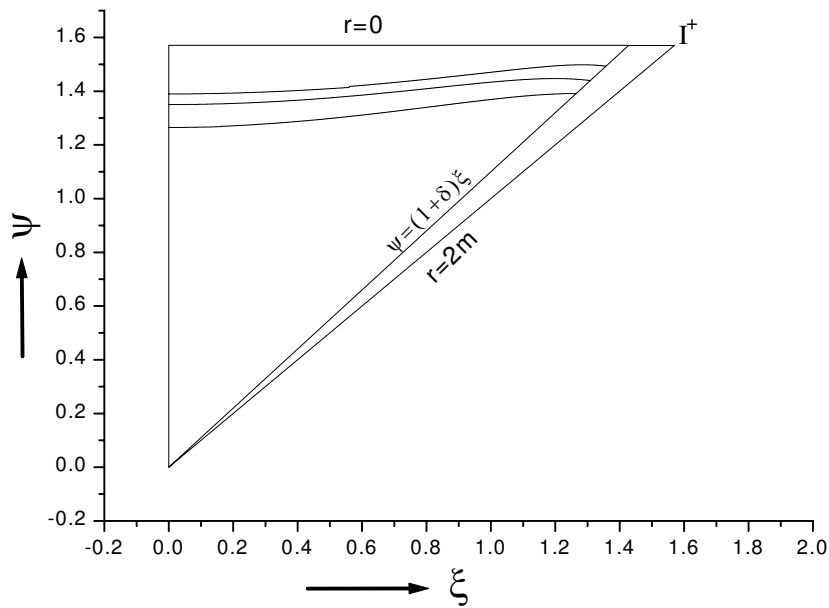}
\end{center}
  \caption{Spacelike hypersurfaces, of constant mean extrinsic curvature, in a
Carter-Penrose diagram of the Schwarzschild geometry are shown. Only
a few typical hypersurfaces are shown corresponding to $K=0.1$,
$0.3$ and $0.8$.}
\end{figure}
\begin{center}
\begin{tabular}{|c|c|c|c|c|}
\hline
$Kr_{s}$ & $r_{i}/r_{s}$ & $\psi _{initial}$ & $\psi _{final}$ & $length$ \\
\hline
10$^{-3}$ & .762 & 1.240 & 1.359 & 3.77 \\ \hline
10$^{-2}$ & .760 & 1.243 & 1.362 & 3.82 \\ \hline
0.1 & .745 & 1.265 & 1.391 & 4.18 \\ \hline
0.2 & .727 & 1.289 & 1.434 & 4.74 \\ \hline
0.3 & .7098 & 1.310 & 1.462 & 5.23 \\ \hline
0.4 & .687 & 1.336 & 1.485 & 5.78 \\ \hline
0.5 & .673 & 1.350 & 1.497 & 6.14 \\ \hline
0.8 & .630 & 1.390 & 1.500 & 6.59 \\ \hline
\end{tabular}
\end{center}
\bigskip

\textbf{Table 1: }Eight $K$ slices for different values of the mean
extrinsic curvature, $K$, are described by the corresponding values for the
initial value of $r$, $r_{i}$, the initial value of $\psi $ in the
Carter-Penrose diagram, $\psi _{initial}$, the final  value, $\psi _{final}$
, and the proper length of the slices.

\section{Summary and Discussion}

It was seen that the generic behaviour of the Schwarzschild singularity is
that it has infinite proper length and occurs at an instant of time in the
appropriate frame. (Of course, one can find frames in which there will be
parts of the singularity that occur \textit{before} and parts that occur
\textit{after} any given point on the singularity.) This fact appears even
in the naive use of the usual Schwarzschild coordinates. It appears more
clearly in the CKS coordinates used for the Carter-Penrose diagram. However,
there is no geometric significance to these coordinates. A frame in which
the singularity \textit{does} occur simultaneously is the one for which we
use the York time, i.e. hypersurfaces of constant mean extrinsic curvature
are the surfaces of simultaneity. It was shown that in this frame the proper
length goes to infinity as the one third power of the York time, which goes
to infinity at the singularity. That the collapse does, indeed, occur in
this frame is shown by the fact that the volume shrinks to zero as the
inverse two thirds power of the York time.

The interesting aspect of this observation is that the simple-minded view of
the spherical collapse to a point needs to be modified. The collapse \textit{
is} to a point in some sense but is to a \textit{line} in another. From
outside the black hole it does indeed seem to collapse to a point. Using the
picture of the black hole provided by the Schwarzschild coordinates inside,
we see that the coefficient of the solid angle does shrink to zero and so
the collapse is spherically symmetric. However, that coefficient is a
timelike parameter. The point is that $r=0$ is not \textit{where} the
collapse occurs but \textit{when} it occurs. A new direction (along the $t$
coordinate) has opened up inside the black hole, in some sense orthogonal to
our usual space, on to which the matter collapses. As such, the volume does
\textit{not} shrink as the cube of $r$ (which is a time) but as a lower
power. The ambiguity of how to treat the asymptotic behaviour of the length,
is taken care of by using the CKS coordinates. It was found that the proper
length goes to infinity as $-\epsilon ^{-1/8}\ln \epsilon $. Here $\epsilon
\rightarrow 0$ as $r\rightarrow 0$. However $r$ depends not only on $
\epsilon $ but also on $\xi $, which measures a place on the singularity. In
CKS coordinates, $r\thicksim 2r_{s}\epsilon ^{1/2}(\sec \xi )^{1/2}$. Thus
we see that $\Delta s\thicksim -r^{1/2}\ln r$ for $\xi \neq \pm \pi /2$.
Consequently, the volume $V\thicksim r^{3/2}\ln r$ for these hypersurfaces,
so long as we stay away from $\xi \neq \pm \pi /2$. The problem here is that
we need to include the points $\xi =\pm \pi /2$. This problem is avoided by
considering York, or $K$-slices. It is found here that $\Delta s\thicksim
K^{1/3}\ln K$ and $V\thicksim K^{-1}\ln K$, unlike the suture model in which
the $\ln K$ is missing. Thus the length goes to infinity even faster than
for the suture model.

It is of interest to also see what happens to the asymptotic behaviour of
the length and the volume for $(n+1)-$dimensions. Generally $\Delta
s\thicksim K^{1/3}\ln K$ still. However, $V\thicksim K^{1-2n/3}\ln K$. Thus
for higher dimensional gravity the singularity becomes stronger. For $(2+1)-$
gravity $V\thicksim K^{-1/3}\ln K$, significantly slower than for the usual $
(3+1)-$gravity, and for $(1+1)-$gravity $V\thicksim K^{1/3}\ln K$. Hence it
\textit{diverges} instead of going to zero! Thus, in $2-$d gravity there
would not be a meaningful \textit{collapse} to a black hole singularity.

\section{References}

1. A. Hussain and A. Qadir, Phys. Rev. \textbf{D63}(2001)083502-1. \newline
2. C. W. Misner, K. S. Thorne and J. A. Wheeler, \textit{Gravitation,}
(Freeman, 1973). \newline
3. A. Pervez, A. Qadir and A.A. Siddiqui, Phys. Rev. \textbf{D51}(1995)4598.
\newline
4. A. Qadir and A.A. Siddiqui, J. Math. Phys. \textbf{40}(1999)5883. \newline
5. A. Qadir and J.A. Wheeler, \textit{From SU(3) to Gravity, }eds. E.
Gotsman and G. Tauber (Cambridge University Press, Cambridge, 1985)
pp.383-394. \newline
6. A. Qadir and J.A. Wheeler, Nuclear Physics \textbf{B} (Proc. Suppl.),
\textit{\ }eds. Y.S. Kim and W.W. Zachary, \textbf{6} (1989) 345-348.
\newline
7. A. Qadir, \textit{Proceedings of the Fifth Marcel Grossmann Meeting},
eds. D.G. Blair and M.J. Buck$_{{}}$ingham, World Scientific, 1989, pp.
593-624. \newline
8. S.W. Hawking and G.F.R. Ellis, \textit{The Large Scale Structure of
Spacetime}, (Cambridge University Press 1973). \newline
9. D.G. Brill and F. Flaherty, Commun. Math. Phys. \textbf{50}(1976)157;
Ann. Inst. Henri Poincare \textbf{28}(1978)335; A. Lichnerowich, J.Math.
Pure Appl. \textbf{23}(1944)37; J.W. York, Phys. Rev. Letters, \textbf{28}
(1972)1082; J.E Marsden and F.J. Tipler, \textit{Maximal Hypersurfaces and
Foliations of Constant Mean Curvature in General Relativity}, Preprint,
Berkeley (1978); L. Smarr and J.W. York, Phys. Rev. \textbf{D17}(1978)2529;
D.M. Eardley and L. Smarr, Phys. Rev. \textbf{D19}(1978)2239.\newline
10. D.R. Brill, J.M. Cavallo and J.A. Isenberg, J.Math. Phys. \textbf{21}
(1980)2789.

\end{document}